\documentstyle[prl,aps]{revtex}
\input epsf.tex

\begin{document}
\draft
\twocolumn[
\widetext

\title{New Incommensurate Elongated-Triangle Phase in Quartz }
\author{P. Saint-Gr\'egoire$^{1,2}$, I. Luk'yanchuk$^{1,3}$, E.Snoeck$^2$,
C.Roucau$^2$, V. Janovec$^{4,5}$}
\address{$^1$G.D.P.C., CC 026, Universit\'e Montpellier II , France}
\address{$^2$CEMES-LOE, BP 4347, 31055, Toulouse, C\'edex, France}
\address{$^3$L.D.Landau Institute for Theoretical Physics, Moscow, Russia }
\address{$^4$University of Technology, 16117, Liberec Czech Republic }
\address{$^5$Institute of Physics, Acad. Sci., 18040 Prague, Czech Republic }

\date{\today }
\maketitle

\widetext
\begin{abstract}
\leftskip 54.8pt
\rightskip 54.8pt

We present for the first time the evidence for a thermodynamically
stable incommensurate elongated-triangle ($ELT$) phase  in quartz  observed
by transmission electron microscopy at structural $\alpha -\beta $
transition. The  phase sequence on cooling is: uniform $\beta $ phase -
stripe incommensurate phase (ferroelastic) -   incommensurate
equilateral-triangle ($EQT$) phase
(ferroelectric) -    incommensurate $ELT$ (ferroelectric and ferroelastic) -
uniform $\alpha $ phase. The  $ELT$ blocks
could be responsible for the large  light
scattering in the vicinity of $\alpha-\beta$ transition.
\end{abstract}


]

\narrowtext

\setlength{\parindent}{5pt} \leftskip -10pt \rightskip 10pt Although the
structural $\alpha $-$\beta $ phase transition in quartz is known since more
than a century, interest has be renewed in the last decades due to discovery
of an incommensurate phase existing between $\alpha $ (low temperature) and $%
\beta $ (high temperature) phases, in a temperature range of approximately 1
K around 847 K (For a review see \cite{Rev}\cite{RevPreprint}). The $\beta $
phase is hexagonal, with the space group $P6_z2_x2_y$, having x and y type
basal plane binary axes. The $\alpha $-$\beta $ transition is induced by a
rotation of $SiO_4$ tetrahedra around x axes by an angle $\eta $ \cite{Grimm}
which reduces symmetry of the $\alpha $ phase to $P3_z2_x$. The main
mechanism of the incommensurate structure formation was shown by Aslanyan
{\it et al.} \cite{AL} to be the strong coupling between the elastic strain
and the spatial gradient of $\eta $ which is responsible for a finite-$q$
instability of $\eta $ at the critical temperature $T_i$ and for a regular
space modulated structure of $\eta \left( r\right) $ between $T_i$ and
lock-in transition temperature $T_c$ below which the system recovers
homogeneity of $\eta $.

Transmission electron microscopy (TEM) observations allow one to follow the
transformation from incommensurate to $\alpha $ phase. They display the
incommensurate state as a regular triangular pattern of equilateral
microdomains with $\pm \eta $\cite{Landuyt} \cite{Amel}\cite{SRS}. This
Equilateral-Triangle ($EQT$ ) phase corresponds to the minimum of Landau
functional calculated in \cite{AL}. Close to $T_c$ this structure is often
broken by appearance of elongated dagger-shaped triangles, pointing in
several directions so that the global organization looks rather chaotic.
{}From this one could suspect \cite{Landuyt} that a new phase attempts to
nucleate in the vicinity of the lock-in transition but large thermal
gradient is perturbing its formation.

In this letter we present the results obtained after improving the
conditions, by minimizing the temperature differences in the sample and also
by performing more slowly the temperature changes during observations. We
report the first TEM observation of the new incommensurate
Elongated-Triangle ($ELT$) phase which is formed in a small temperature
region near $T_c$. This observation, together with the free energy
calculations, gives a new insight into the lock-in phase transition in
quartz. We claim that $ELT$ phase nearby $T_c$ becomes thermodynamically
more stable than $EQT$ phase and ideally the sequence of phases $EQT$ -$ELT$
-$\alpha $ should appear when temperature decreases. Note that several
experiments report on a stripe incommensurate phase in a small temperature
region of $0.1K$ just below $T_i$ (\cite{RevPreprint} and references
therein). We shall not consider this phase here. Finally, we discuss the
macroscopic properties of the $ELT$ phase and relate the $ELT$ blocks with
the optical inhomogeneities which could be responsible for the huge light
scattering at the $\alpha -\beta $ transition.

\begin{figure}[t]
\epsfxsize=5.4truecm
\vspace{-0.2cm}
\centerline{\epsffile {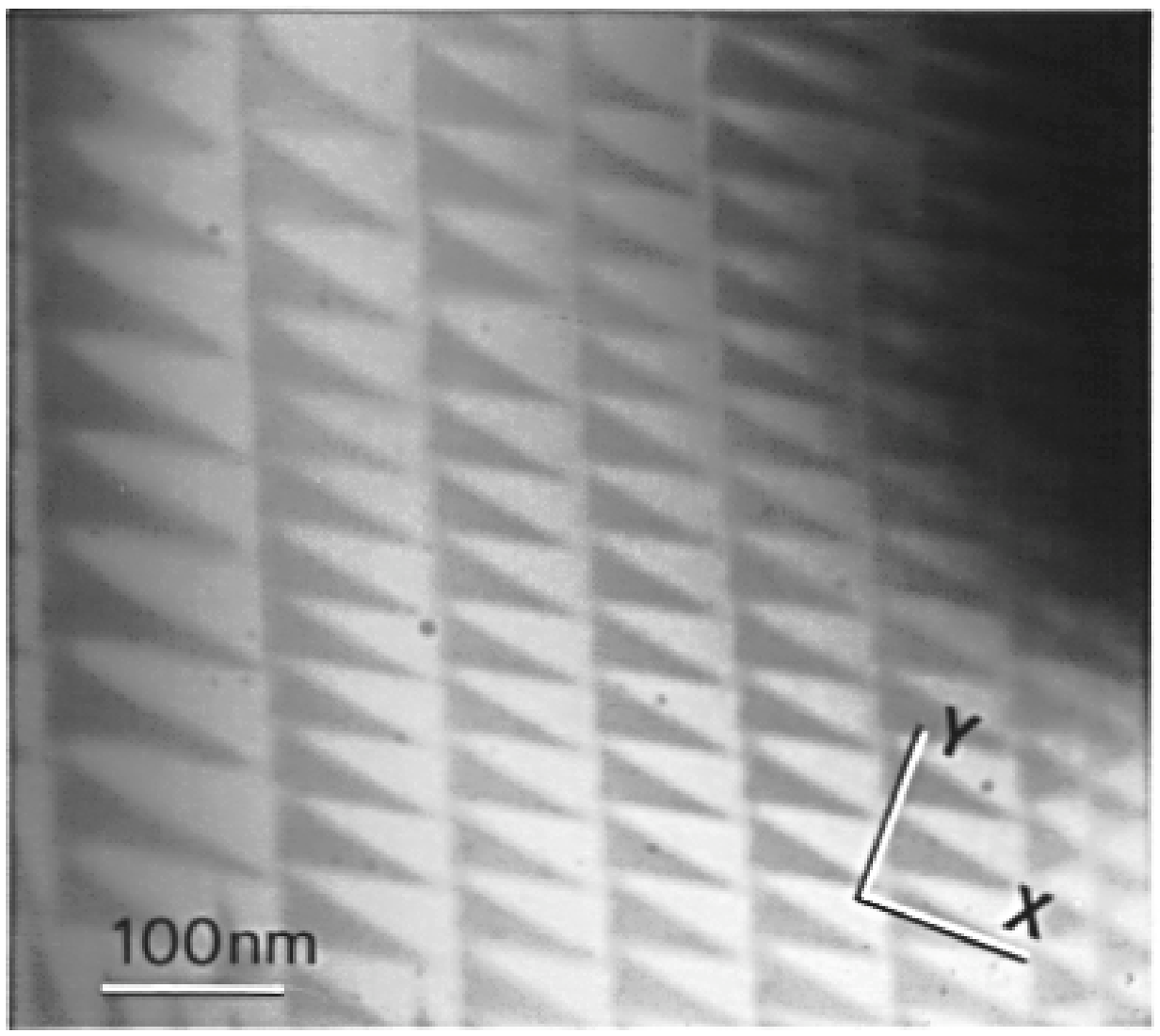}}
\vspace{-2.3cm}
\caption{Dark field image (TEM, Bragg spot (110) ) of $ELT$ phase close to
$T_c$. Note the coexistence  with $EQT$ phase (right bottom corner) and
dislocations in the $ELT$ texture, induced by a slight temperature decrease
from the right to the left.}
\label{fig1}
\end{figure}

In our TEM experiments, similarly as in \cite{SRS}, the specimens were first
mechanically polished until a thickness of approximately 300 $\mu $m was
reached, than further polished with the help of a dimpling device until the
central area reached 20 $\mu $m; finally, an ionic thinning (argon, under a $%
15^0$ angle of incidence) allowed one to obtain samples sufficiently
transparent to electrons. To minimize damages due to the electron beam, the
observations started at temperatures close to the $\alpha -\beta $
transition. Experiments were performed with a JEOL 200 CX electron
microscope.

In these conditions the $ELT$ phase was observed both on cooling and on
heating in a narrow temperature interval about $0.1K$ just above the $\alpha
$ phase. We obtained the micrographs showing regular $ELT$ texture in ranges
of about 1000 nm (Fig.1). In this temperature region the distance between
consecutive parallel walls is of the order of 100 nm. The $EQT$ to $ELT$
phase transition is of the first order and both phases may coexist in the
crystal. It is difficult to determine the mechanism of this change in
general case, however, as already observed in precedent studies \cite{SRS}%
\cite{PSG}, the elongated triangles often appear as a regular formation
within the boundaries between blocks of the $EQT$ phase, at temperatures
close to $T_c$. This suggests that $ELT$ phase could nucleate within $EQT$
block boundaries.

Now, we show that the observed $ELT$ phase does become energetically more
stable than $EQT$ phase near lock-in transition. First we remind that two
different limit approaches are  used to describe incommensurate states. The%
{\it \ Landau functional approach} \cite{AL}, which is valid near $T_i$,
gives a satisfactory explanation of the appearance of the incommensurate
state as a linear combination of  sinusoidal waves. The{\it \ domain wall
approach} \cite{Walker} assumes that the space distribution of the order
parameter acquires a domain-like texture between the uniform domain states  $%
\pm \eta $. This approximation seems to be more relevant nearby the lock-in
transition where the domain-like space distribution of $\eta (r)$ is indeed
observed by TEM, the width of domain walls $\xi $ being substantially
smaller than the distance $h$ between walls.

To examine the lock-in transition region we follow the domain wall approach
considering the incommensurate state as a texture of interacting domain
walls with junctions and intersections between them. The calculations of the
domain texture energy in quartz that we present here, are of the same nature
as those performed by P.Bak {\it et al.} \cite{Bak} in the  study of the
commensurate-incommensurate transition in $2H-TaSe_2$ and in rare-gas layers
adsorbed on graphite.

Three contributions to the energy of a domain texture (calculated with
respect to the uniform $\alpha $ phase) are: (i) The energy of the domain
walls. (ii) Interaction between nonparallel walls which cross in the
vertices; we include to this contribution the energy of the vertices
themselves and call all together the ''vertex energy''. (iii) Interaction
between parallel domain walls.

Consider these contributions in detail.

(i) The energy of isolated domain walls is a function of their orientation
and temperature. According to TEM observations the domain walls in quartz
are parallel to the z axis and only approximately parallel to the x-type
crystallographic axes. Walker concluded \cite{Walker} that the equilibrium
orientation of domain walls is tilted away from the x-type axes by a small
angle $\varepsilon $, which is  of $10^0-15^0$ near $T_c$ and vanishes near $%
T_i$ according to Landau functional calculations \cite{AL}. The reason for
this is that the exact x orientation is not symmetrically prominent because
the point symmetry group $2_z^{\prime }$ (the prime reminds that the
operation changes sign of $\eta $) of a domain wall slightly rotated around
z is the same as for the wall of x-orientation \cite{VJ}.

Fig.2a shows six equivalent equilibrium orientations of domain walls with
tilting angles of $\pm \varepsilon $ for $1^{\pm }$,$2^{\pm }$,$3^{\pm \ }$
walls participating in    domain texture formation. The $EQT$ structure is
formed by the equidistant sets of either $1^{+}$,$2^{+}$,$3^{+}$ or $1^{-}$,$%
2^{-}$,$3^{-}$ domain walls, as shown in Fig.2b. These two degenerate states
form blocks of $EQT$ phase (with typical size of $0.6-1\mu m$) which are
rotated by $\pm \varepsilon $ with respect to the crystallographical x-type
axes. We identify $ELT$ phase with a triangular structure having internal
angles $2\epsilon $, $120^0-2\epsilon $ , $60^0$. Six block states are
formed by  the following six equivalent sets of domain walls: $\left(
1^{+},1^{-},2^{-}\right) $, $\left( 1^{+},1^{-},3^{+}\right) $, $\left(
2^{+},2^{-},3^{-}\right) $, $\left( 2^{+},2^{-},1^{+}\right) $, $\left(
3^{+},3^{-},1^{-}\right) $ and $\left( 3^{+},3^{-},2^{+}\right) $. The $ELT$
phase corresponding to the set $\left( 2^{+},2^{-},3^{-}\right) $ is
sketched in Fig.2c.

\begin{figure}[t]
\vspace{-1cm}
\epsfxsize=5truecm
\centerline{\epsffile {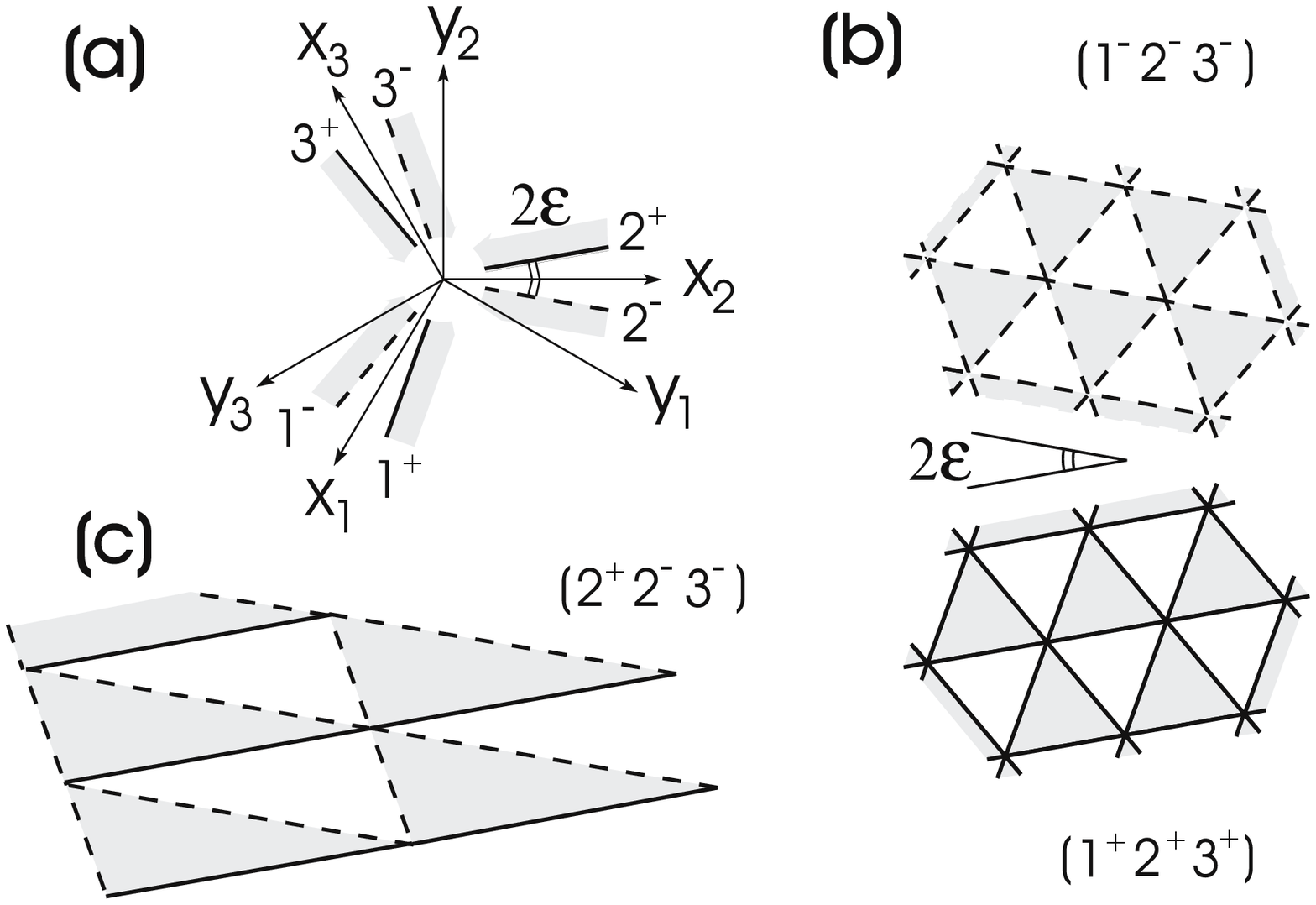}}
\vspace{-2cm}
\caption{Domain walls and textures in quartz; white and grey regions
correspond to the $\pm \eta $ (a) Equivalent orientations of separate domain
wall with respect to crystallographic axes. (b) Equivalent $EQT$ block states
(c) One of the six possible $ELT$ block states}
\label{fig2}
\end{figure}

To account for the lock-in transition we assume that the domain walls energy
is positive in the $\alpha $ phase and reverses its sign above $T_c$ as it
was suggested in \cite{Landuyt}. In the incommensurate state the free energy
is lowered by a packing of a large number of walls into a lattice which is
stabilized by the repulsion between walls and the positive vertex energy.
Below $T_c$ the domain walls disappear from the sample to decrease the
energy.

Consider a domain wall slightly tilted from the equilibrium orientation by a
small angle $\varphi $. We write its energy per unit length near $T_c$ as:

\begin{equation}
e=A\left( T_c-T\right) +G\varphi ^2  \label{Domwall}
\end{equation}
with $G>0$. The increasing of the distances between domain walls near the
lock-in transition, clearly seen in TEM observations is the consequence of
vanishing  the walls energy at $T_c$.

Recall two important properties of domain walls provided by their $%
2_z^{\prime }$ symmetry \cite{WandG}. a) Unlike the uniform non-polar  $%
\alpha $ and $\beta $ phases, walls display the ferroelectric polarization
along z-axis which is antiparallel for the $1^{+},2^{+},3^{+}$ and $%
1^{-},2^{-},3^{-}$ sets. b) Domain walls carry a nonzero elastic strain and
the displacement field undergoes a change $\Delta {\bf u}$ within the wall .

(ii) The vertex energy $Q$ depends on the number and orientation of the
crossing domain walls. Possible vertices have been analyzed in \cite{Walker}
\cite{Ex}. As shown in \cite{WandG} a Burgers vector ${\bf b=}\sum \Delta
{\bf u}_i$ can be associated with a vertex. Vertices with ${\bf b}=0$
represent a line singularity with higher elastic energy than with ${\bf b}=0$.
 The latter condition is fulfilled both for the vertices of $EQT$ and $ELT$
phase. This makes the $EQT$ and $ELT$ structures more favorable
than other incommensurate textures.

(iii) We write the interaction energy between two adjacent parallel domain
walls per unit length as: $Be^{-h/\xi }$, where the distance $h$ between
them,  is assumed to be larger than their width $\xi $.

\begin{figure}[t]
\epsfxsize=5truecm
\centerline{\epsffile {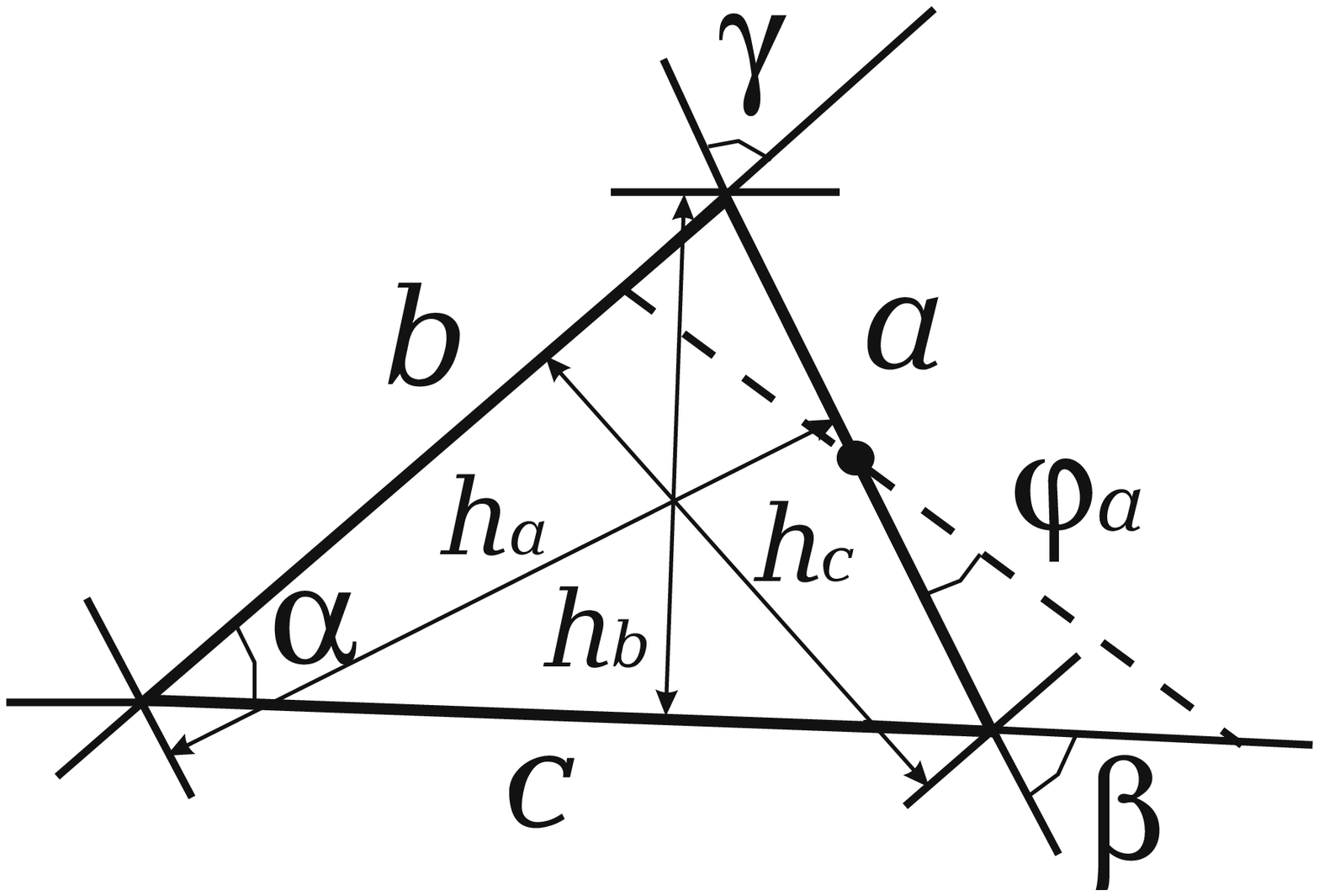}}
\vspace{-3cm}
\caption{Parametrization of triangular domain texture}
\label{fig3}
\end{figure}

To compare the energies of $ELT$ phase and $EQT$ phase quantitatively we
parametrize their geometrical structure by the elementary triangles with
internal angles $\alpha $,$\beta $,$\gamma $, heights $h_a$,$h_b$,$h_c$ and
area $S$ as shown in Fig.3. We examine also the stability of these phases
with respect to the slight tilting of the edges $a$,$b$,$c$ from their
equilibrium orientations (e.g. $2^{+},2^{-},3^{-}$ for $ELT$ phase and $1^{+}
$,$2^{+}$,$3^{+}$ for $EQT$ phase) by angles $\varphi _a$,$\varphi _b$,$%
\varphi _c$. Then $\alpha $,$\beta $,$\gamma $ are presented as: $\alpha
=\alpha _0-\varphi _c+\varphi _b$, $\beta =\beta _0-\varphi _a+\varphi _c$, $%
\gamma =\gamma _0-\varphi _b+\varphi _a$ where $\alpha _0$,$\beta _0$,$%
\gamma _0$ are the angles for the equilibrium orientation that are $60^0$
for $EQT$ phase and respectively $2\varepsilon ,120^0-2\varepsilon ,60^0$
for $ELT$ phase.

In these terms the energy of the domain  texture per unit area is expressed
as:

\begin{eqnarray}
{\cal F} &=&\frac 1S[\frac 12\left( e_aa+e_bb+e_cc\right) +\frac 12
Q\label{Gen} \\
&&+B(ae^{-h_a/\xi }+be^{-h_b/\xi }+ce^{-h_c/\xi })]  \nonumber
\end{eqnarray}
where $Q$ (either $Q_{ELT}$ or $Q_{EQT}$) is the vertex energy. The domain
wall energy per unit length, $e_w$ ($w=a,b,c$), is given by (\ref{Domwall})
with $\varphi =\varphi _w$. One can show that very close to $T_c$ where the
wall density goes to zero the third term in (\ref{Gen}) becomes
exponentially small in comparison with the second one. The most stable
configuration is the result of minimization of (\ref{Gen}). We choose $S$
and $\varphi _a$,$\varphi _b$,$\varphi _c$ as variational parameters, taking
into account that $a=\sqrt{2S}\nu _a$, $h_a=\sqrt{2S}/\nu _a$, where $\nu
_a=\left( \frac{\sin \alpha }{\sin \beta \sin \gamma }\right) ^{1/2}$ and
the analogous expressions for $b$, $h_b$, $\nu _b$, $c$, $h_c$, $\nu _c$.
Neglecting the parallel walls interaction, minimizing ${\cal F}$ over $S$
and expanding the result over the small values of $\varphi _a$,$\varphi _b$,$%
\varphi _c$ we obtain:

\begin{eqnarray}
{\cal F} &=&-[A\left( T_c-T\right) \cdot p+A\left( T_c-T\right) \left( \zeta
_a\varphi _a+\zeta _b\varphi _b+\zeta _c\varphi _c\right)   \nonumber \\
&&+G\left( \varphi _a^2+\varphi _b^2+\varphi _c^2\right) \cdot p]^2\frac 1{2Q%
}  \label{Res}
\end{eqnarray}

where: $p=\nu _a+\nu _b+\nu _c$, $\zeta _a=\frac 12(\nu _c-\nu _b)(\cot
\beta +\cot \gamma )$ $+\frac 12\nu _a(\cot \beta -\cot \gamma )$ and the
analogous expressions for $\zeta _b$, $\zeta _c$ are taken at $\alpha
=\alpha _0$, $\beta =\beta _0$, $\gamma =\gamma _0$. When $T\rightarrow T_c$%
, $S$ diverges as $2Q^2/p^2A^2\left( T_c-T\right) ^2$. Below lock-in
transition the domain texture is absent and ${\cal F}=0$.

For $EQT$ phase $p\simeq 3.2$ and $\zeta _a,\zeta _b,\zeta _c=0$ which
results to the quadratic dependence of energy (\ref{Res}) on the tilting
angles. Therefore the $EQT$ phase is stable against domain walls tilting
from original orientations and conserves its $EQT$ form.
With $\varphi _a,\varphi _b,\varphi _c=0$ we have:

\begin{equation}
{\cal F}_{EQT}\simeq -5.2A^2\left( T_c-T\right) ^2/Q_{EQT}  \label{Triq}
\end{equation}

For $ELT$ phase if one takes $\varepsilon =10^0$ one obtains: $p\simeq 4.9$,
$\zeta _a\simeq -0.28$, $\zeta _b\simeq -3.6$, $\zeta _c\simeq 3.9$.
Minimization of (\ref{Res}) with respect to $\varphi _a$,$\varphi _b$,$%
\varphi _c$ yields: $\varphi _a\simeq -0.035A\left( T-T_c\right) /G$, $%
\varphi _b\simeq -0.44A\left( T-T_c\right) /G$, $\varphi _c\simeq
0.47A\left( T-T_c\right) /G$ and

\begin{equation}
{\cal F}_{ELT}\simeq -8.2A^2\left( T_c-T\right) ^2/Q_{ELT}  \label{ETP}
\end{equation}

When temperature increases the domain walls tilt from equilibrium
orientation at $T_c$ in order to reduce the smallest angle $\alpha =\alpha
_0-\varphi _c+\varphi _b$ as: $2\varepsilon -0.9A\left( T-T_c\right) /G$.
This suggests that $ELT$ phase approaches the stripe  phase when temperature
goes to $T_i$.

Comparing (\ref{Triq}) and (\ref{ETP}) we get that ${\cal F}_{ELT}<{\cal F}%
_{EQT}$ if the energies of $EQT$ and $ELT$ vertices are not very different,
or, more precisely, if the condition $Q_{EQT}/Q_{ELT}>0.6$ is satisfied
which is likely since both types of vertices are elastically nondefective
and their energies are expected to be of the same order. Then, close to $T_c$%
, the $ELT$ phase appears to be more stable than the $EQT$ phase.

The above conclusion is valid only very close to the lock-in transition,
where the interaction between parallel walls is negligible because of the
large distance $h$ between them. The adjacent walls interaction becomes
important at higher temperatures where $h$ diminishes. Assuming now that in
this temperature region the third term in (\ref{Gen}) is dominating and
neglecting the vertex energy we obtain that the $EQT$ phase is the most
stable one as the state with the maximal wall concentration when the
distance between them is fixed. This is compatible with Landau functional
calculations \cite{AL} which gives the stability of $EQT$ phase nearby $T_i$%
. This qualitative consideration shows that at some critical temperature $%
T^{*}$, $T_c<T^{*}<T_i$, the first order phase transition between $ELT$ and $%
EQT$ phase is expected.

Compare now the macroscopic electric and elastic properties of $EQT$ and $ELT
$ phases. Since $EQT$ phase is formed by walls carrying an electric
polarization along z (either $+P_z$ or $-P_z$) it exhibits the macroscopic
{\it ferroelectricity} with the opposite direction of $P_z$ for $1^{+}$,$%
2^{+}$,$3^{+}$ and $1^{-}$,$2^{-}$,$3^{-}$ blocks which can be identified
with ferroelectric domains. Although the domain walls have also a nonzero
elastic strain, it compensates{\it \ }in average for equilateral triangular
texture of domain walls. On the phenomenological level, these properties
\cite{WandG} follow from the $6_z^{\prime }$ point symmetry of $EQT$ phase
\cite{PSG}. In contrast, the $2_z^{\prime }$ point symmetry of $ELT$ phase
is compatible with both the $z-$directed ferroelectricity and the basal
plane spontaneous strain of the crystal. The $ELT$ phase is formed by the $+$
and $-$ domain walls (carrying opposite polarizations along z) which have a
different density so that the macroscopic polarization as well as the
resulting elastic strain does not vanish and blocks can be identified with
ferroelastic and ferroelectric domains. There are several examples of
ferroelastic incommensurate phases\cite{Ferr}, but the $ELT$ phase in quartz
is the first example of incommensurate state which is both {\it \
ferroelastic} and {\it ferroelectric}.

The appearance of $ELT$ phase sheds light on the old problem of the
anomalous strong light scattering at $\alpha -\beta \,$ transition in quartz
\cite{Yakovlev}\cite{Shapiro} which, as it was shown by several authors \cite
{LightScat}, is caused by static columnar optical inhomogeneities of cross
section of $\sim 20\mu m$ which appear in a small temperature interval of $%
\sim 0.1K$ in the region of $\alpha -\beta $ transition. These
inhomogeneities cannot be associated with $\pm P_z$ $EQT$ blocks which
possess the same optical indicatrix because of their $6_z^{\prime }$
symmetry. In contrast, the $ELT\,$ferroelastic blocks have optical
indicatrices of different orientation which results in the spatial
inhomogeneity of the refraction index of the crystal. We propose therefore
that this inhomogeneity is the principal source of the huge light scattering
in quartz.

In conclusion, we have demonstrated the stability of the incommensurate $ELT$
phase just above lock-in transition in a small temperature interval $\sim
0.1K$. Therefore the $\alpha -\beta $ transition in quartz exhibits on
cooling a sequence of three incommensurate phases: the stripe phase
(ferroelastic), the $EQT$ phase (ferroelectric), and the $ELT$ phase
(ferroelectric and ferroelastic). The blocks of $ELT$ phase could be in the
origin of anomalous  light scattering at the $\alpha -\beta $ transition$.$

G.D.P.C. is URA CNRS $n^o$233, CEMES-LOE is UPR CNRS $n^o$8011. The work of
I.L. was supported by regional administration of the Province Languedoc,
France (28PAST8016) by the International Science Foundation and the Russian
Governement, Grant $n^o$MGI300.


\end{document}